\documentclass{kluwer}    
\usepackage{graphicx} \newdisplay{guess}{Conjecture}

\begin{document}
\begin{article}
\begin{opening}         
\title{Scale Interactions and Galaxy Evolution} 
\author{Daniel \surname{Pfenniger}}  
\runningauthor{Daniel Pfenniger}
\runningtitle{Scale Interaction and Galaxy Evolution}
\institute{University of Geneva, Geneva Observatory}
\date{September 30, 2002}

\begin{abstract}
  To understand galaxies and their evolution, it is necessary to
  describe how the different scales interact: how the microscopic
  physics, such as star formation, or the large scale physics, such as
  galaxy interactions may modify the galaxy global shapes.  The
  purpose of this review is to point out some general or recent topics
  related to such scale interactions, both observational and
  theoretical, which are relevant in the present understanding of
  galaxies.
\end{abstract}

\end{opening}  

\section{Introduction}
The purpose of this talk is to review the general mechanisms acting
between the scales in gravitating systems such as galaxies, and that
lead to evolution.  The Universe is certainly hierarchically
organized, and at different scales different phenomena interact with
the adjacent scales.  For the sake of simplicity it has been customary
for a long time to consider one scale at a time and to neglect scale
interactions.  Scientists have been inclined to discard scale
interactions already because distinct disciplines are specialized
according to the astrophysical object sizes.  As a consequence some
attempts to introduce scale interactions such as galaxy mergers met
much resistance.

But nowadays it becomes clearer that as stars are built upon nuclear
and atomic physics, galaxies are also built upon the ISM physics and
star formation processes.  Conversely galaxy clusters and large scale
structures do matter for understanding galaxy evolution, which in turn
cannot be ignored when trying to explain star formation, while stellar
evolution do certainly plays a role in the chemical composition of
matter.  Despite a disparity of phenomena, we see common phenomena
through the scales because some of the involved physics is scale-free,
namely gravitational dynamics.

Statistical physics provides an example how adjacent scales may be
sometimes absorbed by a proper theoretical frame.  The small scale
physics is absorbed by statistical concepts, such as the ``molecular
chaos'', while the large scale physics is absorbed by proper boundary
conditions, such as the extensivity of the system.  Thus one can
obtain a useful description of gases without including the detailed
knowledge of molecules, which are actually quite complex objects.
Without any particle computer simulations statistical physics can
describe the fate of a gas enclosed in a box, the shape of which is
unimportant.

However, if the box size increases too much we know well that a
sufficiently large volume of gas becomes self-gravitating, and then
the usual tools of statistical mechanics do not work well with
gravitating systems.  We briefly review on the reasons in Sect.~2.  In
Sect.~3 we discuss the small scale interactions in galaxies, i.e., the
ISM and star formation physics for which very interesting developments
have occurred in the recent years.  In Sect.~4 some rarely discussed
issues about the large scale interactions of galaxies are presented.
Finally, we conclude in Sect.~5.

\section{Statistical mechanics of gravitating systems}
The fundamental problem of adapting statistical mechanics to
gravitating systems and other non-extensive systems remains to be
worked out.  In fact, statistical mechanics is deeply designed for
\textit{extensive} systems because, first, these systems are very
common in terrestrial conditions, and, second, the extensivity (that
certain macroscopic quantities such as energy and entropy
\textit{scale linearly with the volume}) allows drastic
simplifications of the system description.  It is therefore
paradoxical that concepts like ``temperature'' or ``pressure" are
constantly used in astronomy, but these concepts have been designed
during the XIX$^\textrm{th}$ century especially for extensive systems
in which self-gravitation is negligible.
 
This issue is important because self-gravity is ubiquitous in
astrophysical systems.  With proper statistical mechanics tools we
should be able to understand on a deeper theoretical basis the stellar
and galactic systems, and also the behaviour of N-body simulations.
After all classical statistical mechanics allows to describe particle
systems such as gases without requiring N-body simulations.  The mere
necessity of N-body simulations in astrophysics illustrates the
practical non-applicability of classical statistical mechanics to
systems with long range forces.
    
A basic unsettled issue is the meaning and definition of entropy.  As
long as entropy remains a fuzzy concept (a measure alternatively of
the available phase space, of disorder, or of information) little
change to the situation has to be expected.  However, the awareness is
growing that the extensivity requirement must be abandoned in order to
handle gravitating systems, as well as small $N$ systems (Gross 2002).
For both types of systems the particle interactions play a global
role, therefore the system is not extensive.

The gravitational perfect gas sphere is perhaps the simplest
system where gravitation is combined with statistical concepts, since
it already contains the particular difficulties brought by the long
range gravitational force.  It has been worked out by a number of
authors (e.g.~Ebert 1955; Lynden-Bell \& Wood 1968; Chavanis 2002).
The most notable feature is that below a critical temperature, the
\textit{specific capacity} (or \textit{specific heat}) of the gas
sphere becomes negative.  Such states, forbidden by classical
statistical mechanics, are thermally and dynamically unstable, and are
closely related to Jeans' unstable states.  In other non-gravitational
systems, such negative specific capacity states are called phase
transitions (Lynden-Bell 1999), and are known to develop spontaneously
long range correlations, ``giant fluctuations'', fractal states, etc.
For such states the fluctuations develop until the small scale and
large scale physics react.  At intermediate scales, new scaling
relations may occur.  The system, at least in a transient phase, may
develop scaling relations different from the classically assumed
extensivity.

In other words, in negative specific capacity states scale interactions
play a decisive role, as well as the boundary conditions.  The small
scale physics is often the fastest, so is the most important to include
faithfully.  For negative specific capacity states it is necessary to
include a description of the effective microscopic physics, contrary
to the positive heat capacity states where the small scale molecular
physics is irrelevant.

\begin{figure} 
\hspace{0pc}\includegraphics[scale=0.535,angle=90]{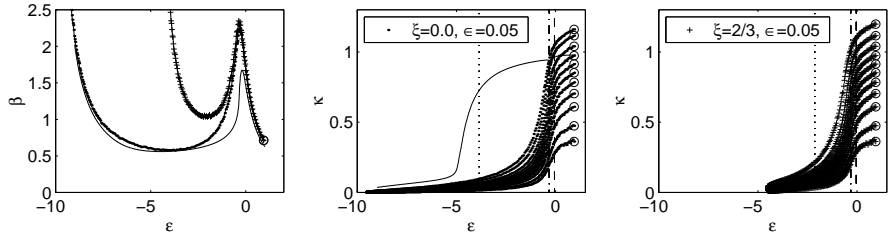}
\caption[]{\textbf{Left:} Energy $\varepsilon$ as 
  a function of the inverse temperature $\beta$ of a particle system
  softened with a Plummer potential with softening $\epsilon=0.05$,
  and confined by a spherical box of unit radius. The thin solid curve
  corresponds to the analytical model of Follana \& Laliena (2000),
  and the thick dots to a similarly softened N-body simulation (Huber
  \& Pfenniger 2002), while the crosses follow another model with a
  small scale repulsive softening.  Negative specific capacity phases
  occur when the slope of these curves is positive, so are narrower
  when the particles
  have a repulsive core.  \\
  \textbf{Middle:} The Lagrangian radii $\kappa$ of the Plummer
  softenend particle system do not follow the analytical prediction
  during the negative specific capacity phase, marked by vertical
  dashed lines, during which long range correlations in
  phase space develop.  \\
  \textbf{Right:} The Lagrangian radii of the repulsive softening core
  system.  } \vspace{0pc}
\label{SIMUL}
\end{figure}

Such effects have been recently studied by Huber \& Pfenniger (2002)
via N-body simulations.  In this study, the gas sphere is simulated by
gravitating particles slowly dissipating energy.  As long as the
system specific capacity is positive, the particle system follows well
the analytical theory of Plummer softened particles developed by
Follana \& Laliena (2000), but as soon as the specific capacity
becomes negative, i.e., the system is gravitationally unstable, strong
differences occur (Fig.~\ref{SIMUL}).  In such latter states the
growth of long range correlations in phase space, fractal states,
occurs in the particle system which are not descriptible with the
analytical theory, but resemble the correlations of Larson (1981)
observed in the cold ISM.  The specific behaviour and evolution
followed by the N-body system is then strongly dependent on the
adopted small scale physics, i.e., the particular softening
properties.

All these theoretical issues have important consequences for the way
we understand galaxies and other gravitating systems:
\begin{enumerate}
\item In N-body simulations, during the fragmenting phases developing
  down to the particle level one cannot trust the numerical model as
  simulating faithfully the real natural systems as long as the
  microscopic physics included in the model does not match the
  relevant microscopic physics in the natural system.  In particular,
  negative specific capacity at the particle level is up to now never
  included in N-body simulations.  In the future this aspect should be
  taken into considerations since in galaxies and cosmology such
  gravitationally unstable states, i.e., with negative specific
  capacity, occur frequently.
  
  As found in Huber \& Pfenniger (2002) the bias in today's particle
  simulations is that softening acts systematically as keeping the
  specific capacity at the particle level at positive values.  The
  same bias occurs in gas simulations where the small scale physics is
  generally represented by a perfect gas.
  
\item The important consequence for real galaxies is that the
  fragmenting and clustering phases must depend strongly on the small
  scale physics.  As for stars, the general properties of galaxies
  should be searched also and perhaps principally in their internal
  properties, and less in their initial conditions, as has been for a
  long time exclusively considered.  The problem is hard because it
  involves extremely non-linear and chaotic phenomena, such as the
  star formation process, which must be understood first if we want to
  model their impact at the galaxy level.

\end{enumerate} 

\section{Internal scale interactions in galaxies}
\subsection{The new stellar and galactic dynamics}
The ``microscopic'' physics of galaxies, i.e., the ISM physics, the
physics of star formation, the physics associated with stellar
evolution is crucial in order to understand galaxies.  It is obvious
that a mass condensation that is called galaxy must contain a minimum
amount of stars, so the star formation process belongs by definition
to the galaxy formation process.  For a long time this early epoch of
galaxy formation was thought to be confined to a restricted time
period, so it was convenient to simplify the description of galaxies
as collisionless ensemble of stars.  With this model N-body
simulations could account for a large number of galaxy properties, and
the tendency was to neglect the ISM and stellar physics.  Later
numerous simulation works did include some representation of gas
dynamics, as well as recipes of star formation.

But a fundamental aspect, perhaps much more important for galactic and
stellar dynamics that 2-body relaxation or Liouville's theorem, the
overall mass loss from stars, has been considered only very recently
(e.g., Kroupa, 2002; Bournaud \& Combes 2002).  Indeed, besides
energy, stellar populations do return a substantial fraction ($>20\%$)
of their mass to the ISM over $5-10\, $Gyr, especially after the red
giant phase.  As consequence momentum, angular momentum and mechanical
energy mixing must also occur between stellar populations and the
diffuse, energy dissipative hot gas component.  This is especially
true for elliptical galaxies, which were considered for a long time as
perfect gasless and collisionless pure stellar dynamical systems!\@
Since the mixing of several percents of mass due to stellar mass loss
means a substantial dissipation for a mechanical system, it turns out
that over time-scales longer than a couple of Gyr galaxies must be
viewed as essentially dissipative structures.  In comparison to the
effect of stellar mass loss, the much longer 2-body relaxation time is
completely irrelevant concerning the effective global relaxation.
Over time-scales shorter than a few Gyr, still an order of magnitude
larger than the galaxy dynamical time, the collisionless description
remains however valid.

As example of consequence that stellar mass loss may lead to, Bournaud
\& Combes (2002) show that several phases of bar destruction and
reformation are possible in N-body simulations including mass loss and
external gas accretion.

\subsection{The new ISM physics}
From observational constraints molecular clouds appear increasingly as
transient structures with a lifetime reduced by an order of magnitude
with respect to earlier estimates, stars form essentially over a
free-fall time according to Elmegreen (2000).

The more diffuse HI also is subject to an deep rediscussion about its
lifetime and origins. For a long time HI in the outer galactic disks
was seen as almost primordial, or at least a long lived phase, but
recent observational works suggest something very different.  The
study of M101 by Smith et al. (2000) suggests that the HI is a
by-product of the FUV radiation, HI is observed in proportion of the
exciting UV radiation, which means that a cold molecular susbtrat must
exist even in the outer HI disks.  Deep photometric observations by
Cuillandre et al.~(2001) of the extreme outer disk of M31 reveal the
unambiguous existence of stars, while extinction of stars and
background galaxies reveal that dust exists in the HI. Blue clustered
stars indicate that stars do form inconspicuously even in the extreme
outer regions of M31's disk.  But as far as we know stars require cold
and dense molecular clouds for forming, thus there is little freedom
left from the conclusion that molecular hydrogen does exist even in
extreme outer galactic disks (Allen 2001).  Similar conclusions were
reached from distinct motivations for explaining the baryonic dark
matter in spirals (Pfenniger \& Combes 1994).

\subsection{Internal energetics and global galaxy parameters}
Traditionally the global galaxy parameters have been explained as
relics of particular \textit{initial conditions}.  The general
correlations among galaxies, such as the Hubble sequence, and the
Tully-Fisher relation, have been sought as relics of correlations
existing in the initial conditions.

However the virial theorem in the form $E_\mathrm{tot} = -
E_\mathrm{kinetic}$ shows immediately that to condense matter with
little specific energy from infinity into a bound system, one must
dissipate the present specific kinetic energy.  The dissipative nature
of galaxy formation is essential.  So a galaxy rotating at
$V_\mathrm{rot} =200\,\mathrm{km\,s^{-1}}$ must have dissipated away
its specific kinetic energy $V_\mathrm{rot}^2$, at least about
$208\,\mathrm{eV/nucleon}$, a value intermediate between typical
chemical binding energy (of order of eV) and nuclear binding energy
(of order of MeV).  In addition the galaxy mass is made of a fraction
of stars, each nucleon in a Sun-like star requires additionally to
have dissipated away its present thermal energy ($\approx
400\,\mathrm{eV/nucleon}$).  In comparison, a black-hole requires
about $2 \cdot 10^6$ times more specific dissipation ($0.5\, c^2
\approx 1\,\mathrm{GeV/nucleon}$), which means that the fractional
mass of black-holes becomes dominant in the global energetics when it
exceeds $\approx 0.5 \cdot 10^{-6}$.  Already the modest Milky Way
black-hole with $2-3\cdot 10^6 \,M_\odot$ exceeds this threshold.

Internal processes are powerful enough to control or modify the
process of galaxy formation or transformation, provided that the
emitted energy can interact with the galaxy scale.  At any time what
matters is the exchanged \textit{power} between the scales.  The
maximum dynamical power $L_\mathrm{dyn}$ that a gravitating system can
exchange is given by the ratio of its energy and dynamical time, which
takes a particularly simple form in term of virial velocity $V$
(cf.~Pfenniger 1991):
\begin{equation}
L_\mathrm{dyn} = \frac{|E|}{t_\mathrm{dyn}} =\frac{V^5}{G}
\end{equation}
This order of magnitude scaling is resembling the IR Tully-Fisher
scaling ($L_\mathrm{IR} \propto V_\mathrm{rot}^{4.9}$) not only in the
exponent, but also in the zero point.  For example if $V =
200\,\mathrm{km\,s^{-1}}$ then $L_\mathrm{dyn} = 1.2\cdot 10^{10} \,
L_\odot$.  The physical interpretation of this is that independently
of the origin of the Tully-Fisher relation, galaxies deliver in the
form of light, i.e., nuclear energy, a power comparable to the power
required to mechanically transform secularly its global structure.

Thus the secular impact of radiation over the global galaxy mechanical
energy appears important, especially because galaxies are nowadays
known to be only semi-transparent.  The similarity of values and
velocity scaling between luminosities and dynamical power is unlikely
a coincidence.  Via global dynamical reaction (bars and spirals in
marginally stable disks), spiral galaxies can adjust their size to the
internal energy production related to stellar activity (SN explosions,
WR winds, outflows, radiation, HI holes, etc.).  The galaxy global
parameters are then not only determined by the initial conditions of
formation, but much more by the internal micro-physics (ISM and star
formation).
 
\section{External scale interactions in galaxies}
\subsection{The energy--angular momentum budget of interactions}
Galaxies frequently interact with the higher scale, with gas infall,
accretion, tidal interactions, galaxy collisions or mergers.  But
often the infall, accretion or merger processes are viewed as simple
mass addition processes.  Consequently, the history of galaxy build-up
is summarized by a ``merger tree'', which leads to the approximate
view that galaxy formation is a hierarchical \textit{addition} of
masses as a function of time.  In fact, N-body simulations show that
mergers may be fairly complex, and until a quasi steady state of the
merger remnant is found, much occur, not only regarding the
morphological evolution of the primary galaxy, but also because each
important event ejects a non negligible fraction of mass, sometimes
over 10\%, escapes to sufficiently large distances to be considered as
infinity (i.e., the ejected mass return time-scale is much longer than
the merger remnant dynamical time).
 
Mass escaping the system means that angular momentum and energy is
also transported with values not necessarily equal to the merger
remnant ones.  When considering galaxies, to first order they are well
explained by the scale-free Newtonian gravitational physics, so the
total mass is irrelevant; one can thus normalize with the mass.  A
succession of merger events correspond then to a complicated walk in
the \textit{specific} angular momentum--energy parameter space.
  
It is easy to see with a simple dimensional model how angular momentum
constraints the galaxy global parameters.  The system specific energy
$e$ and the specific angular momentum $l$ read:
\begin{equation}
e = \frac{1}{2}\left( V^2_\mathrm{rot}+\sigma^2\right) - 
\frac{GM}{\alpha R}\,,
\quad 
\beta l = \alpha R \, V_\mathrm{rot}\,,
\end{equation}
where the kinetic energy part is decomposed into the ordered bulk
rotational part with the average rotational velocity $V_\mathrm{rot}$,
and the disordered kinetic energy velocity dispersion $\sigma$.  $M$
is the total mass and $R$ the virial radius.  $\alpha$ and $\beta$ are
constants of order of 1.  By the virial theorem, the system specific
energy is also minus the kinetic energy:
\begin{equation}
e = -\frac{1}{2}\left( V^2_\mathrm{rot}+\sigma^2\right)  < 0 \,.
\end{equation}
Now it is useful to express the previous equations for the latter
fragile but observable quantities $R$, $V_\mathrm{rot}$, and $\sigma$
as functions of the robust but hardly observable quantities $M$, $l$,
and $e<0$:
\begin{equation}
\alpha R=  \frac {GM}{2|e|}, \quad 
V_\mathrm{rot}= 2|e| \frac {\beta l}{GM}, \quad
\sigma^2=2|e| \left[1 -  2 |e| \left(\frac { \beta l }{ G M} \right)^2 \right].
\end{equation} 
\begin{figure} 
  \hspace{1.3pc}\includegraphics[scale=0.55,angle=0]{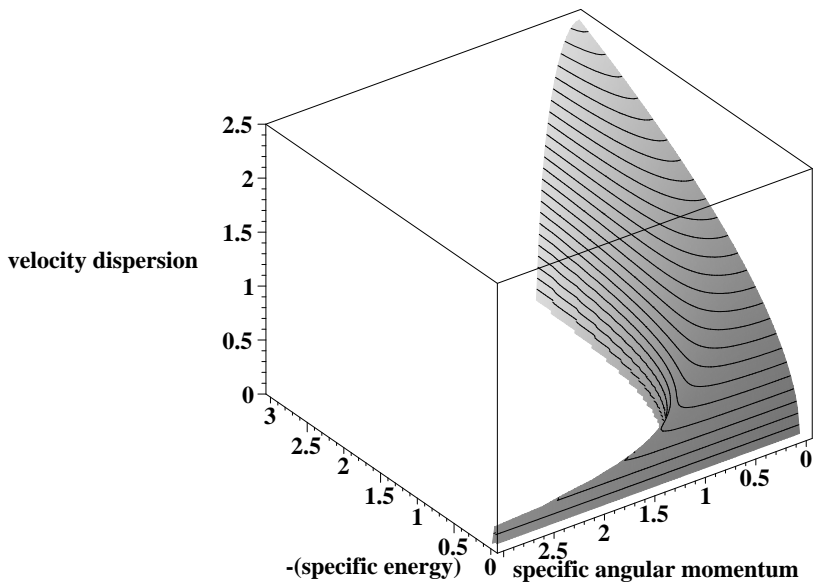}
\caption[]{Velocity dispersion as function of specific angular momentum and
  energy. } \vspace{0.5pc}
\label{SIGMA}
\end{figure}

This simple model captures the essential effects due to interactions
or other perturbations to the systems integral quantities.  The system
radius decreases if $M$ decreases or $|e|$ increases (i.e., when
energy is lost), but is insensitive to variations of $l$.  The
rotational velocity $V_\mathrm{rot}$ increases either if energy is
lost, or if $l$ increases, or if $M$ decreases.  Finally, the velocity
dispersion $\sigma$ has a richer behaviour, summarized by
Fig.~\ref{SIGMA}, with a large forbidden region: $\sigma$ decreases,
i.e., the system cools, when $l$ is sufficiently increased.  Low $l$,
large $|e|$, and large $\sigma$ correspond to hot, slowly rotating
systems, while large $l$, small $|e|$, and small $\sigma$ correspond
to disk systems.

With this gained insight it is now easy to predict conditions
favorable for interactions or other factors leading not necessarily
toward hotter, slower rotating systems, as usually expected, but
toward the opposite, disky systems: one factor is to raise the system
specific angular momentum, and the other is to inject also energy, to
decrease $|e|$.  So prograde mergers, and internal energy injection by
stellar activity both favors a galaxy toward a disky state.
    
\subsection{HVC accretion and dark matter halo shape}
The angular momentum constraint can be used in the scenario that High
Velocity Cloud (HVC) infall secularly increases the Galaxy mass by a
few $M_{\odot}\,\mathrm{yr^{-1}}$ (e.g., Blitz et al. 1999).  First,
the Magellanic Stream as a source of HI has the wrong angular momentum
vector since it is on almost polar orbit.  Second, since the HVC's
appear to contain individually an order of magnitude more dark matter,
as the HI outer disk does, one must conclude that the accreted dark
matter is substantial, and contains a similar specific angular
momentum as the baryons.  Consequently the dark mass should be
expected to adopt a shape of a fast rotator (outwards flaring thick
disk), which are known to depart substantially from the spheroidal
shape almost exclusively contemplated in the literature for pressure
supported systems.  Notice that a substantial rotational support for
the dark matter may resolve several current conflicts about cuspy CDM
cores, since ``hollow'' dark mass distributions surrounding maximum
optical disks are then possible (Pfenniger \& Combes 1994).

\section{Conclusions}
Scale interactions are important in astrophysics because the negative
specific capacity states are ubiquitous.  These states are unstable
and develop phase-space correlations which trigger the interactions of
scales.  This poses several deep problems for the current way to
perform N-body simulations, but this also indicates that galaxies
should not be seen as isolated system insensitive to their scale
boundary conditions.  At small scale stellar mass loss is secularly
important, over several Gyr galaxies must be seen as dissipative
structures.  Also the stellar energy input is important enough to
compete with dynamics and is sufficient to determine the global spiral
parameters.  At large scales, galaxy interactions lead to a a walk
through the mass, energy, and angular momentum space, since at each
interaction a substantial amount of matter may be recycled through the
IGM.  Also the angular momentum constraint linked to dark matter rich
gas infall indicates that flaring disk-like rotationally supported
dark matter distributions should be considered too, not exclusively
pressure supported spheroids.

\acknowledgements 
\vspace{-0.5pc}
\begin{footnotesize}
  The conference organizers and Kiel University staff are gratefully
  acknowledged for the inordinate amount of efforts dedicated for
  insuring a smooth and enjoyable conference.  This work has been
  supported by the Swiss National Science Foundation.
\end{footnotesize}
\theendnotes

\end{article}

\begin{thebibliography}{}
  
\bibitem[\protect\citeauthoryear{Allen}{2001}]{Allen} Allen, R.J.
  \newblock {2001}, \newblock {in Gas and Galaxy Evolution, ASP Conf.
    Proc.}, 240, 331

\bibitem[\protect\citeauthoryear{Blitz et al.}{1999}]{BlitzEtal}
  Blitz, L., Spergel, D.N., Teuben, P.J., et al.  \newblock {1999},
  \newblock {\em Astrophys. J.}, 514, 818--843
  
\bibitem[\protect\citeauthoryear{Bournaud \&
    Combes}{2002}]{BournaudCombes} Bournaud, F., Combes, F.  \newblock
  {2002}, \newblock {\em Astron. Astrophys.}, 392, 83--102
  
\bibitem[\protect\citeauthoryear{Chavanis}{2002}]{Chavanis} Chavanis,
  P.H.  \newblock {2002}, \newblock {\em Astron. Astrophys.}, 381,
  340--356
  
\bibitem[\protect\citeauthoryear{Cuillandre et
    al.}{2001}]{Cuillandre_etal} Cuillandre, J.-C., Lequeux, J.,
  Allen, R.J., et al.  \newblock {2001}, \newblock {\em Astrophys.
    J.}, 554, 190--201
  
\bibitem[\protect\citeauthoryear{Ebert}{1955}]{Ebert} Ebert, R.
  \newblock {1957}, \newblock {\em Zeitschrift f\"ur Astrophysik}, 37,
  217
  
\bibitem[\protect\citeauthoryear{Elmegreen}{2000}]{Elmegreen}
  Elmegreen, B.G.  \newblock {2000}, \newblock {\em Astrophys. J.},
  530, 277--281
  
\bibitem[\protect\citeauthoryear{Follana \&
    Laliena}{2000}]{FollanaLaliena} Follana, E., Laliena, V.
  \newblock {2000}, \newblock {\em Physical Review E}, 61, 6270--6277
  
\bibitem[\protect\citeauthoryear{Gross}{2002}]{Gross} Gross D. H. E.
  \newblock {2002}, \newblock {\em Physica A}, 305, 99--105
  
\bibitem[\protect\citeauthoryear{Huber
    \&Pfenniger}{2002}]{HuberPfenniger} Huber, D., Pfenniger, D.
  \newblock {2002}, \newblock {\em Astron. Astrophys.}, 386, 359--378
  
\bibitem[\protect\citeauthoryear{Kroupa}{2002}]{Kroupa} Kroupa, P.
  \newblock {2002}, \newblock {Habilitation Thesis, University of
    Kiel}
  
\bibitem[\protect\citeauthoryear{Larson 1981}{1981}]{Larson} Larson,
  R.B.  \newblock {1981}, \newblock {\em Monthly Not. Royal Astron.
    Soc.}, 194, 809--826
  
\bibitem[\protect\citeauthoryear{Lynden-Bell}{1999}]{Lynden-Bell}
  Lynden-Bell, D.  \newblock {1999}, \newblock {\em Physica A,} 263,
  293--304
  
\bibitem[\protect\citeauthoryear{Lynden-Bell \&
    Wood}{1968}]{Lynden-BellWood} Lynden-Bell, D., Wood, R.  \newblock
  {1968}, \newblock {\em Monthly Not. Royal Astronom. Soc.} 138, 495
  
\bibitem[\protect\citeauthoryear{Pfenniger}{1991}]{Pfenniger}
  Pfenniger, D.  \newblock {1991}, \newblock {in Dynamics of Disc
    Galaxies, G\"oteborg Univ., B. Sundelius ed.,} p. 389
  
\bibitem[\protect\citeauthoryear{Pfenniger \&
    Combes}{1994}]{PfennigerCombes} Pfenniger, D., Combes, F.
  \newblock {1994}, \newblock {\em Astron. Astrophys.}, 285, 94--118
  
\bibitem[\protect\citeauthoryear{Smith et al.}{2000}]{Smithetal}
  Smith, D.A., Allen, R.J., Bohlin, R.C., et al.  \newblock {2000},
  \newblock {\em Astrophys. J.}, 538, 608--622

\end{thebibliography}
\end{document}